\documentclass{emulateapj}

\newcommand{\msunperyr}{\ensuremath{\mathrm{M}_{\sun}\,\mbox{yr}^{-1}}}

\newcommand{\eqref}[1]{equation (\ref{#1})}

\newcommand{\sub}[1]{\ensuremath{_{\mbox{\scriptsize#1}}}}

\begin{document}
\slugcomment{To Appear in The Astrophysical Journal}
\title{On the Likelihood of Planet Formation in Close Binaries}
\shorttitle{ON THE LIKELIHOOD OF PLANET FORMATION IN CLOSE BINARIES}
\author{Hannah Jang-Condell}
\shortauthors{JANG-CONDELL}
\affil{University of Wyoming, Department of Physics \& Astronomy\\
1000 E.~University, Dept 3905, Laramie, WY 82071}
\email{hjangcon@uwyo.edu}

\begin{abstract}
To date, several exoplanets have been discovered 
orbiting stars with close binary companions ($a\lesssim30$ AU).  
The fact that planets can form in these dynamically challenging 
environments implies that planet formation must be a robust process.  
The initial protoplanetary disks in these systems from which 
planets must form should be tidally truncated to radii of a few 
AU, which indicates that the efficiency of planet formation must be 
high.  Here, we examine the truncation of circumstellar 
protoplanetary disks in close binary systems, 
studying how the likelihood of planet formation is affected 
over a range of disk parameters.  
If the semimajor axis of the binary is too small  
or its eccentricity is too high, the disk will have too little 
mass for planet formation to occur.  However, we find that 
the stars in the binary systems known to have planets should have once 
hosted circumstellar disks that were capable of supporting planet formation 
despite their truncation. 
We present a way to characterize the feasibility of planet formation 
based on binary orbital parameters such as stellar mass, companion mass, 
eccentricity and semi-major axis.  Using this measure, we can 
quantify the robustness of planet formation in close binaries and 
better understand the overall efficiency of planet formation in general.  
\end{abstract}

\keywords{
planets: formation ---
protoplanetary disks ---
binaries: close
}

\section{Introduction}

Close binary systems are a challenging environment for planet
formation because a binary companion will dynamically perturb the
protoplanetary disk.  
However, several exoplanets have been found in 
S-type orbits (circumstellar) around stars in
binary systems with semi-major axes of less than 30 AU 
\citep[e.g.][]{2006Raghavan_etal}.  
These systems represent interesting test cases for planet formation 
theories.  By comparing the frequency of planets in 
binary systems with that around single stars, we can learn about 
how robust the planet formation process can be.  
In this paper, we address the topic of circumstellar planets
(S-type orbits), as opposed to circumbinary planets (P-type orbits).  

HD 188753 is a hierarchical triple system, with 
a 12.3 AU separation and orbital eccentricty of 0.5 
between the A and BC components.
\citet{HD188753} claimed the detection of planet around the A component.  
However, we found that the binary parameters for HD 188753A 
do not allow for in situ planet formation 
to occur \citep[][henceforth Paper I]{HJChd188753}.  
Indeed, it was later found that there was no planet there after all
\citep{2007Eggenberger_etal}. 

On the other hand, $\gamma$ Cep has a separation of 20 AU 
and eccentricity of 0.4, and has a well-established 
gas giant planet orbiting the A component at $\sim$2 AU
\citep{2003Hatzes_etal,2011Endl_etal}.  
Although at first glance, its orbit is not very different 
from that of HD 188753, 
we found that the binary parameters of $\gamma$ Cep A do allow 
in situ planet formation to occur
\citep[][henceforth Paper II]{2008HJCMugrauerSchmidt}.

\begin{table*}[bth]
\caption{
\label{tab:binaries}
Binary systems analyzed in this paper.  
}
\begin{tabular}{lcccccc}
& HD 188753A & $\gamma$Cep A & HD 41004A & HD 41004B & HD 196885A & $\alpha$Cen B\\
\hline\hline
$M_*$ ($M_{\odot}$) & 1.06 & 1.40   & 0.7     & 0.4     & 1.31   & 0.93\\
$M_b$ ($M_{\odot}$) & 1.63 & 0.41   & 0.4     & 0.7     & 0.45   & 1.1 \\
$\mu$              & 0.39 & 0.77   & 0.64    & 0.36    & 0.74   & 0.46\\ 
$a$ (AU)         & 12.3 & 20.2   & 20      & 20      & 21     & 23.5\\
$e$              & 0.5  & 0.41   & 0.4     & 0.4     & 0.42   & 0.52\\
$m_p\sin i$ ($M_J$)& ---  & 1.85   & 2.5     & 18.4    & 2.96   & 0.0034\\
$a_p$ (AU)         & ---  & 2.1    & 0.006 &$7\times10^{-4}$& 2.6 & 0.04 \\
$e_p$              & ---  & 0.05   & 0.7     & 0.08    & 0.46   & 0 \\
reference          & a    & b      & b,c     & d       & e      & f \\
$N_M$	           & 9    & 12     & 12      & 12      & 12     & 12 \\
$N\sub{DI}$        & 0    & 1      & 1       & 1       & 1      & 1 \\
$N\sub{CA}$        & 0    & 8      & 7       & 4       & 8      & 3 
\end{tabular}
\tablecomments{Orbital parameters without subscripts ($a$ and $e$) 
refer to those of the binary.  
The subscript `p' refers to planetary or substellar companions.  
}
\tablerefs{
(a) \citet{2007Eggenberger_etal}, 
(b) \citet{2011Endl_etal}, 
(c) \citet{2004Zucker_HD41004},
(d) \citet{2003Zucker_HD41004},
(e) \citet{2008Correia_etal}, and
(f) \citet{2012Dumusque_etal}
}

\end{table*}

Circumstellar planets have been discovered in a few other binary systems 
of comparable separations.  These include 
HD 196885 \citep{2008Correia_etal,2009Fischer_etal,2011Chauvin_etal}, 
$\alpha$ Cen \citep{2002Pourbaix_etal,2012Dumusque_etal,2013Hatzes}, and 
HD 41004.  Interestingly, 
HD 41004 has substellar companions in both A and B components: 
HD 41004 Ab is a 2.5 $M_J$ planet \citep{2004Zucker_HD41004}
and HD 41004 Bb is a brown dwarf \citep{2003Zucker_HD41004}. 
The binary orbits and planet characteristics of the 
systems discussed here are listed in Table \ref{tab:binaries}.  

One way to study the robustness of planet formation in binaries is 
to study the long-term dynamical stability of those planets 
\citet[e.g.][]{1999HolmanWiegert}.  However, the focus here 
in this work is on the formation of planets in the first place.  
Rather than doing a detailed calculation of the dynamics of planet 
formation as in, for example, \citet{2011Thebault}, 
we want to understand planet formation more broadly by considering 
the effects of binary interactions with the initial protoplanetary disk.  
That is, given that the protoplanetary disk will be truncated 
by a binary companion, what is the likelihood that planet 
formation can take place at all?

Calculations of disk truncation by binary companions show that 
stars in binary systems are still able to retain 
circumstellar disks \citep{1994ArtymowiczLubow}.  
This is supported by 
millimeter and submillimeter observations of pre-main sequence 
binaries which show that binary stars with separations of 
tens of AU are able to retain their disks 
\citep{1996Jensen+,1995OsterlohBeckwith,1994Jensen+,1990Beckwith+}.  

The objective of this paper is to give a measure for the difficulty 
of planet formation across a range of binary star parameters.  
We consider S-type orbits, in which the planet orbits one of the stars 
in the binary system, as opposed to P-type orbits, which refer to 
circumbinary planets.  
We examine the initial protoplanetary disk around a given star and 
calculate the degree of truncation of this disk.  
Rather than determining where exactly planets should form in 
these truncated disks, we focus on whether planet formation 
is possible at all.  

In this paper, we extend the methods of determining the feasibility of
planet formation in close binaries as developed in Papers I and II
to a wider range of binary star configurations.
For a given set of binary parameters, we develop a formalism for 
describing the feasibility of 
planet formation as measured by the number of disk models that 
would allow for either core accretion or disk instability to occur.

\section{Methods}

\subsection{Truncated Disk Models}

We base our estimates of the likelihood of planet formation 
for a given binary system on 
the range of disk parameters that permit enough mass in the 
disk for planet formation to occur, given that a circumstellar 
protoplanetary disk will be truncated by a binary companion.  
These methods are explained fully in Papers I and II, and 
are summarized briefly here.  

For convenience, we refer to the planet-hosting star as the primary and the 
binary stellar companion as the secondary, 
regardless of the relative masses between the binary components.  
Accordingly, $M_*$ and $M_b$ are the masses of the primary and 
secondary, respectively.  The mass ratio of the binary is defined as 
$\mu = M_*/(M_*+M_b)$.  The binary orbit is defined by 
the semimajor axis $a$, and eccentricity $e$.  

The primary mass, $M_*$, affects both the gravitational potential of
the disk as well as the amount of stellar irradiation of the disk
since luminosity varies strongly with stellar mass.  
We consider a range of masses for the 
primary star from 0.5 to 2.0 $M_{\odot}$.
To calculate the luminosity of the primary star, needed as an input 
parameter for the disk models, we use pre-main sequence models 
for stars of 1 Myr of age \citep{siess_etal}.  The values for 
$T\sub{eff}$, $R_*$ and $L_*$ for each value of $M_*$ selected
are listed in Table \ref{tab:stars}.

\begin{table}
\begin{center}
\caption{\label{tab:stars}Stellar parameters as a function of stellar mass}
\begin{tabular}{ccccc}
Stellar mass & $T\sub{eff}$ & $R_*$ & $L_*$\tablenotemark{$\dagger$} \\
($M_{\odot}$) & (K) & ($R_{\odot}$) & ($L_{\odot}$) \\
\hline\hline
  0.5 &    3770 &   2.1 &   0.8 \\
  0.7 &    4026 &   2.4 &   1.4 \\
  0.9 &    4209 &   2.5 &   1.8 \\
  1.0 &    4282 &   2.6 &   2.0 \\
  1.1 &    4343 &   2.7 &   2.3 \\
  1.3 &    4447 &   2.9 &   3.0 \\
  1.5 &    4536 &   3.1 &   3.7 \\
  1.7 &    4616 &   3.3 &   4.4 \\
  1.9 &    4690 &   3.4 &   5.0 \\
  2.0 &    4721 &   3.5 &   5.5 
\end{tabular}
\\
$\dagger$Derived from $T\sub{eff}$ and $R_*$.

\end{center}
\end{table}

The two additional parameters that determine the properties of a disk 
are the mass accretion rate ($\dot{M}$) 
and viscosity parameter ($\alpha$).  
For each stellar mass, we calculate a suite of disk models 
by varying $\alpha$ and $\dot{M}$, as
\begin{equation}
\alpha \in \{ 0.001, 0.01, 0.1\}
\end{equation}
and 
\begin{equation}
\dot{M} \in \{ 10^{-9}, 10^{-8}, 10^{-7}, 10^{-6}, 10^{-5}, 10^{-4}\}
\,\msunperyr
\end{equation}
for a total of 18 disk models for each 
set of binary parameters ($M_*,\mu,a,e$).    
The lower accretion rates reflect 
values typically observed in T Tauri disks.
The highest accretion rates are more typical of active FU
Orionis-like objects or very young protoplanetary disks than of
passive T Tauri disks.  Nevertheless, we include these
high values of accretion for completeness.

A given value of $\alpha$ sets the viscosity, according to 
\begin{equation}
\nu = \alpha c_s H
\end{equation}
where $c_s$ is the thermal sound speed and $H$ is the scale height 
of the disk \citep{shaksun}.  These quantities are related by 
$H = c_s\Omega$ where $\Omega$ is the Keplerian angular speed, 
and $c_s = \sqrt{kT/\bar{\mu}}$ where $k$ is the Boltzmann constant, 
$T$ is the midplane temperature, and $\bar{\mu}$ is the mean molecular 
weight of the gas, whose assumed composition is that of molecular 
hydrogen.  The viscosity sets the rate of the mass flow through 
an accretion disk, according to 
\begin{equation}\label{eq:diskprofile}
\nu\Sigma = \frac{\dot{M}}{3\pi}\left(1-r/R_*\right)
\end{equation}
where $\Sigma$ is the surface density, $r$ is the stellocentric radius, 
and $R_*$ is the stellar radius
\citep[see, e.g.,][]{pringle}.  
If we assume that the accretion rate, $\dot{M}$, is constant 
with $r$, then Eq.~\ref{eq:diskprofile} gives a relation 
between $T$ and $\Sigma$ in the disk.  Assuming that the 
only heating sources in the disk are viscous accretion and stellar 
illumination, we calculate $T(r)$ and $\Sigma(r)$ self-consistently.  
Thus, $\alpha$ and $\dot{M}$ can be selected to ``tune'' the 
mass of the disk.  

The truncation radius for each of the 18 disk models for a selected
primary mass is determined from values calculated in
\citet{1994ArtymowiczLubow}, who define the truncation radius to be
where viscous torques balance the tidal torques.  This truncation
radius depends both on the 
Reynolds number, which is controlled by the viscosity $\nu$,
and the binary orbit, represented by the mass ratio
and eccentricity.  That is, for each set of $\dot{M}$ and $\alpha$ for
the disk, and $M_*$, $\mu$, $a$, and $e$ for the binary, there is a
unique disk truncation radius.

\begin{figure}
\plotone{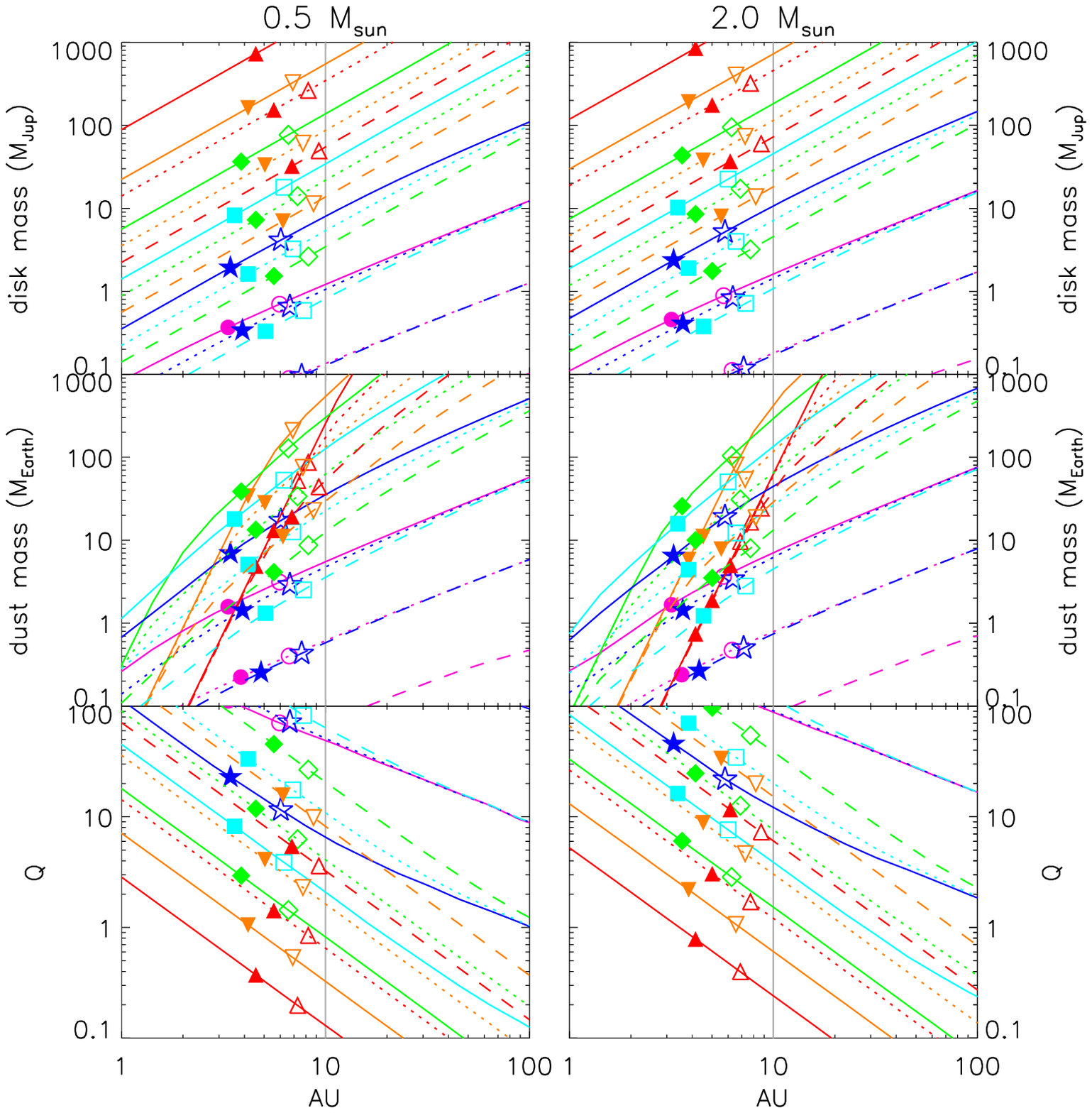}\\[-3ex]
\plotone{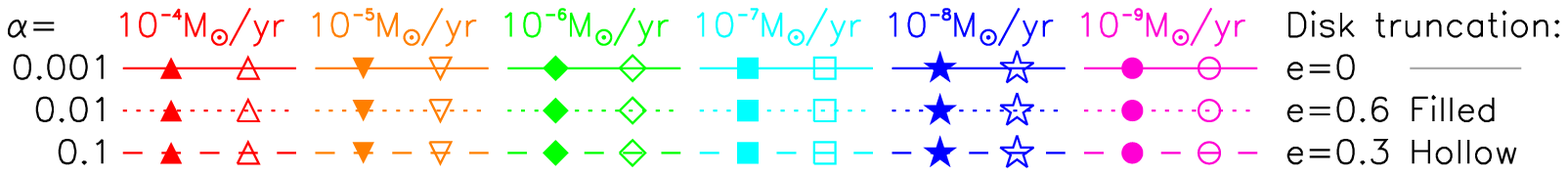}
\caption{\label{fig:diskstruct}
Disk profiles for the least massive (0.5 $M_{\odot}$, left) and 
most massive (2.0 $M_{\odot}$, right) planet host stars examined 
in this work.  
We show the total disk mass (top), total dust mass (middle), 
and Toomre $Q$ parameter (bottom) versus radius for the 
disk parameters explored.  The accretion rate ($\dot{M}$) is 
indicated by line color, while values of the viscosity 
parameter of $\alpha=0.001,0.01,$ and $0.1$ 
are indicated by solid, dotted, and dashed lines.  
The truncation radius for each disk for an equal-mass 
binary companion ($\mu=0.5$) at $a=30$ AU is shown by 
the vertical gray line for $e=0$, by hollow symbols for $e=0.3$
and filled symbols for $e=0.6$.
\\[4ex]
}
\end{figure}

In Figure \ref{fig:diskstruct}, we show how the structure of the disks 
vary with input parameters.  The two sets of plots show disk profiles for 
the least massive (0.5 $M_{\odot}$) and most massive (2.0 $M_{\odot}$)
stars explored.  
Each line represents a selected value of $\dot{M}$ and $\alpha$ 
for a given disk model, as indicated by color and line type 
according to the legend.  For each disk model, we show 
the enclosed disk mass (top), total dust mass (middle), 
and Toomre $Q$ parameter (bottom) as a function of radius.  
The Toomre $Q$ parameter is given by
\begin{equation}\label{eq:Q}
Q = \frac{c_s \kappa} {\pi G \Sigma}, 
\end{equation}
where $\kappa$ is the epicyclic frequency and 
$G$ is the gravitational constant.  Since the disk orbital velocities 
are close to Keplerian, we assume that $\kappa\approx\Omega$.

If the binary orbit is circular, then all disks will be truncated 
at the same radius.  The truncation radius for an equal-mass
binary companion ($\mu=0.5$) at 30 AU in a circular orbit is 10 AU, 
as indicated by a gray line in Figure \ref{fig:diskstruct}.  
However, if the binary companion is eccentric, the truncation 
radius varies with disk properties.  This is because the Reynolds number 
is different in each disk, therefore the balance of tidal and 
viscous torques occurs at different radii.
The truncation radii for a binary orbit with $\mu=0.5$ and $a=30$ AU
with $e=0.3$ $(0.6)$ is shown by hollow (filled) symbols in 
Figure \ref{fig:diskstruct}.  As the eccentricity increases, 
the truncation radii move inward as expected, since the 
periastron occurs at smaller radius with increasing eccentricity.  

The disk profiles shown in Figure \ref{fig:diskstruct} 
vary more with $\dot{M}$ and $\alpha$ than with stellar mass.  
As the stellar mass increases, the disks become slightly more massive 
overall, but with less dust and higher values of $Q$.  
This is because 
the increased stellar illumination heats the disks to higher temperature,
which affects both disk sublimation and pressure support of the gas.  

The Toomre $Q$ parameter decreases with radius in most models 
of protoplanetary disks as a general rule.  
In the disk models presented, the midplane temperature and 
suface density typically behave as $T_c\propto r^{-0.6}$ and 
$\Sigma\propto  r^{-0.9}$, respectively.  
Since $\Omega\propto r^{-3/2}$, the overall variation of 
$Q$ is to decrease with $r$, 

\subsection{Core Accretion vs.~Disk Instability}

Given a set of parameters defining a particular binary system 
($M_*$, $\mu$, $a$, and $e$), we can count 
how many of the 18 truncated disk models constructed around 
the primary star  
(1) have sufficient
total mass to form planets, (2) have sufficient solid material to form
a planet core, or (3) have low enough value of $Q$ for disk instability
to occur.  
If either condition (2) or (3) is satisfied, then (1) is also satisfied, 
so we henceforth consider only the more stringent criteria for 
planet formation based on core accretion and disk instability, 
rather than the total disk mass.  
We define the number of disks in which core accretion is possible 
based on the mass of solid material to be $N\sub{CA}$ 
and the number of disks in which disk instability is possible 
based on the value of $Q$ to be $N\sub{DI}$.

In the core accretion process of planet formation, giant planets 
are formed by first creating a rocky core from the agglomeration of 
dust particles in the disk.  When it reaches sufficient mass, typically 
around 10 $M_{\oplus}$, it becomes massive enough to accrete and 
retain a gaseous atmosphere, and gas giant planet is born.  
Our criterion for determining that a given truncated disk model 
is capable of forming giant planets by core accretion, then, 
is that it contains at least 10 $M_{\oplus}$ of solid
materials, i.e.~dust and ice.  
If the total amount of dust in a given truncated disk model is above 
10 $M_{\earth}$, then we say that that disk is capable of 
supporting planet formation by core accretion.  So, $N\sub{CA}$ is 
the total number of truncated disk models that can support 
planet formation by core accretion for a given set of binary 
parameters.  For example, $N\sub{CA}$ for 
$M_*=0.5\,M_{\odot}$, $\mu=0.5$, $a=30$ AU, and $e=0$
can be read off the dust mass plot in Figure \ref{fig:diskstruct} 
(left center) by counting the number of profiles that intersect the 
truncation radius (gray line) at 10 $M_{\oplus}$ or above, 
giving $N\sub{CA}=12$.  When the eccentricity is raised to $e=0.3$, 
the solid content of the disk model with 
$\dot{M}=10^{-6}\,\msunperyr$ and $\alpha=0.1$ 
drops below 10 $M_{\oplus}$, so $N\sub{CA}=11$.  

The criterion for planet formation by disk instability, on the other 
hand, is determined by the value of the Toomre instability parameter, 
$Q$.  When $Q<1$, the disk is gravitationally unstable
and can rapidly form gas giant planets through fragmentation.  
If a given truncated disk model has $Q<1$ at any point in the disk, 
then we declare that the disk is capable of forming planets by disk 
instability, and $N\sub{DI}$ is the total number of truncated 
disk models that are gravitationally unstable for a 
given set of binary parameters.  
For $M_*=0.5\,M_{\odot}$, $\mu=0.5$, $a=30$ AU, and $e=0$, 
$N\sub{DI}=4,$ and when the eccentricity is raised to $e=0.3$ 
$N\sub{DI}=3$.  

\section{Results and Analysis}

In Papers I and II, we used a fixed set of parameters for $M_*$, $a$,
$e$, and $\mu$ and used those parameters to calculate $N\sub{CA}$, and
$N\sub{DI}$.  Here, we extend the ranges of $M_*$, $a$, $e$, and $\mu$ in
order to see how $N\sub{CA}$ and $N\sub{DI}$ vary with each parameter.
In Table \ref{tab:binaries}, we show our calculations for 
$N\sub{CA}$ and $N\sub{DI}$ for the binary systems 
listed there.  

Our previous results for HD 188753A (Paper I) and 
$\gamma$ Cep A (Paper II) are shown in the first two columns, 
respectively.  We also repeat our calculations for the close binary 
systems HD 41004, HD 196885A, and $\alpha$ Cen B\@.  
Although HD 41004Bb is a brown dwarf rather than a planet, we include 
that system in this paper as a comparison.  

HD 188753A, where there is no confirmed planet, is the only system 
with $N\sub{DI}=0$; all other systems 
have $N\sub{DI}=1$.  Similarly, HD 188753A is the only system 
examined here with $N\sub{CA}=0$.  If we use $N\sub{CA}$ as a proxy 
for the difficulty of planet formation, then $\alpha$ Cen B should be 
the least likely system to harbor a planet, and $\gamma$ Cep A and 
HD 196885A the most likely.  

\begin{figure*}
\includegraphics[width=\columnwidth]{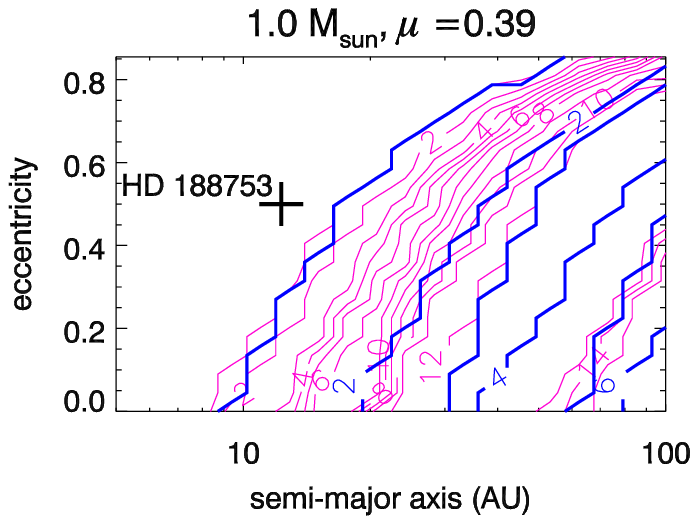}
\includegraphics[width=\columnwidth]{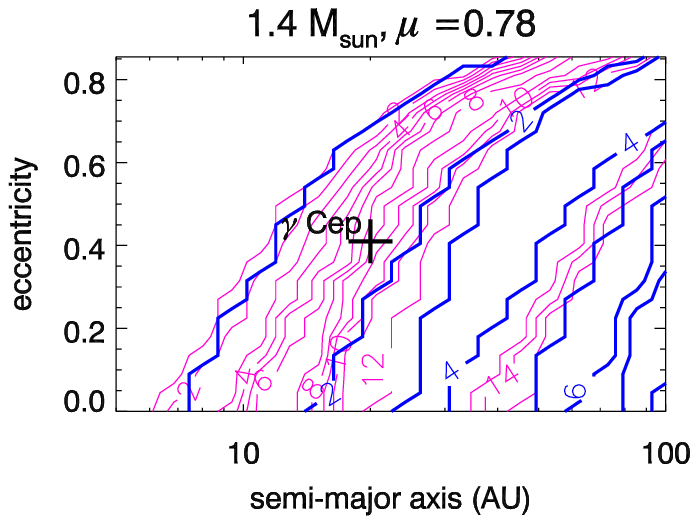}
\caption{\label{fig:indiv}
Planet formation likelihood plots for HD 188753A (left) and 
$\gamma$ Cep A (right). 
Each plot shows how $N\sub{CA}$ and $N\sub{DI}$ vary with 
eccentricity and semi-major axis for the binary orbit for 
a fixed primary and secondary mass.  
Solid magenta contours show values for $N\sub{CA}$ and 
dashed blue contours show those for $N\sub{DI}$.  
The positions of HD 188753A (left) and $\gamma$ Cep A (right)
in $a$-$e$ space are indicated by crosses.  
}
\end{figure*}

In Figure \ref{fig:indiv} we show examples of 
how $N\sub{CA}$ and $N\sub{DI}$ vary with $a$ and $e$, holding $M_*$
and $\mu$ fixed.  That is, the masses of the binary system are held
fixed and their orbits are allowed to vary.  One plot shows where $\gamma$
Cep A falls in parameter space, while the other shows the location of
HD 188753A.
As eccentricity decreases and semi-major axis increases, 
$N\sub{CA}$ (green solid contours) 
and $N\sub{DI}$ (red dashed contours)
both increase, as expected.  
This is because as periastron increases,
the disk truncation radius also increases, allowing a larger 
disk to remain around the star.  $N\sub{DI}$ increases also because
$Q$ tends to decrease with increasing distance from the star, 
and $N\sub{CA}$ increases because more cold disk material remains 
in the disk.

The irregularity in the spacing of the green contours ($N\sub{CA}$),
particularly at large $a$ and small $e$, results from the finite 
sampling of disk models.  If we had included more values of $\dot{M}$ and 
$\alpha$, then there would be more contours in this area. 
Some of the contours bunch up at small separations and low eccentricities, 
largely because the solid content of the disks are not a 
simple function of $\dot{M}$ and $\alpha$, as shown in the middle 
panels of Figure \ref{fig:diskstruct}.  The dust mass versus radius 
profiles of the disks cross each other at small separations, with steeper
profiles at higher accretion rates.  This is because holding $\alpha$ 
constant, higher accretion rates give higher disk mass according to 
Equation (\ref{eq:diskprofile}), but the heating from this accretion causes 
sublimation of the refractory species.  These competing effects 
mean that at large disk radii, the total solid 
content should overall be higher in the more massive, higher accretion 
rate disks, but at small radii, the refractory species sublimate
and those disks have lower solid content.

The values of $N\sub{DI}$ and $N\sub{CA}$ can be fit to a 
polynomial function dependent on 
$a$, $e$, $\mu$, and $M_*$.  The general form of this polynomial is
\begin{equation}
\label{coeffs}
N = \sum_{i,j,k,l} c_{ijkl} \left(\frac{a}{1\mbox{ AU}}\right)^i
e^j \mu^k \left(\frac{M_*}{M_{\odot}}\right)^l.
\end{equation}
Since we want to keep the analytic function as simple as possible, we 
do not try to fit the bunched contours, but rather try to find a 
smooth fit that will give a reasonable approximation for the 
true values of $N\sub{CA}$ and $N\sub{DI}$.  
We find best-fitting polynomial coefficients, $c_{ijkl}$, 
for $N\sub{CA}$, using those points where 
they have tabulated values where $0<N\sub{CA}\leq10$, 
because this avoids the plateau in $N\sub{CA}$ values 
at larger separations and low eccentricity.  
For $N\sub{DI}$, we use tabulated values where
$0<N\sub{DI}\leq5$.  As shown in Figure \ref{fig:indiv}, 
values of $N\sub{DI}$ higher than 5 occur only at 
$a\gtrsim50$ AU and at low eccentricity, and in a region 
where $N\sub{CA}>10$.  

We find that for $N\sub{CA}$ we get reasonably good fits 
for a function linear in $a$ and $e$ and quadratic in $\mu$ and $M_*$.  
For $N\sub{DI}$, the fitting function is linear in all four parameters.  
The best-fitting coefficients are tabulated in Table \ref{tab:coeffs}.
Comparing the directly calculated and tabulated values for $N\sub{CA}$ 
to the polynomial fit for all tabulated values where 
$N\sub{CA}\leq10$, the root-mean-square 
deviation is 0.62.  For $N\sub{DI}\leq5$, the rms 
deviation is 0.44.  

The values of $N\sub{CA}$ and $N\sub{DI}$ for the binary 
systems considered in this paper as calculated using 
this polynomial equation are tabulated in 
Table \ref{tab:checkstars}.
Since both $N\sub{CA}$ and $N\sub{DI}$ cannot be less than $0$, 
negative values should be interpreted as $0$.  
The resulting values are all within $\pm1$ for $N\sub{CA}$, 
and within $\pm0.6$ for $N\sub{DI}$ 
of the directly calculated values shown in Table \ref{tab:binaries}.  
The polynomial values are more discrepant for larger values of $N\sub{CA}$, 
underestimating the true values.  This is a result of the polynomial 
equation smoothing over the bunching of $N\sub{CA}$ contours, 
as discussed above.  

\begin{table*}
\begin{center}
\caption{\label{tab:coeffs} Coefficients for 
Equation (\ref{coeffs}). 
}
\begin{tabular}{llrrrr}
  &  & $i=0, j=0$  & $i=1, j=0$  & $i=0, j=1$  & $i=1, j=1$ \\ \hline\hline
\multicolumn{6}{c}{Core Accretion} \\ \hline\hline
$k=0$, & $l=0$ 
 & $ -6.7585$& $  0.4308$& $  2.9369$& $ -0.4411$\\
$k=1$, & $l=0$ 
 & $ -2.0956$& $  1.6243$& $ -9.5050$& $ -1.4578$\\
$k=2$, & $l=0$ 
 & $  1.8452$& $ -0.5589$& $ 16.0528$& $  0.3879$\\
$k=0$, & $l=1$ 
 & $  1.5486$& $ -0.1609$& $ -1.9967$& $  0.1696$\\
$k=1$, & $l=1$ 
 & $ -9.3964$& $  0.2216$& $  3.5221$& $ -0.3374$\\
$k=2$, & $l=1$ 
 & $  7.9117$& $ -0.3244$& $ -5.4470$& $  0.4971$\\
$k=0$, & $l=2$ 
 & $ -0.2951$& $  0.0364$& $  0.4372$& $ -0.0363$\\
$k=1$, & $l=2$ 
 & $  2.7651$& $ -0.0993$& $ -1.6003$& $  0.1275$\\
$k=2$, & $l=2$ 
 & $ -2.2214$& $  0.0920$& $  2.1346$& $ -0.1406$\\
\hline\hline
\multicolumn{6}{c}{Disk Instability} \\ \hline\hline
$k=0$, & $l=0$ 
 & $  0.3921$& $  0.0521$& $ -0.8329$& $ -0.0431$\\
$k=1$, & $l=0$ 
 & $ -0.1135$& $  0.1106$& $  1.1926$& $ -0.0975$\\
$k=0$, & $l=1$ 
 & $ -0.2704$& $ -0.0066$& $  0.3131$& $  0.0012$\\
$k=1$, & $l=1$ 
 & $  0.2944$& $ -0.0296$& $ -0.7153$& $  0.0327$\\
\hline\hline
\end{tabular}

\end{center}
\end{table*}

\begin{table*}
\begin{center}
\caption{\label{tab:checkstars} 
Calculated values of $N\sub{CA}$ and $N\sub{DI}$ from
Equation (\ref{coeffs}).
}
\begin{tabular}{ccccccc}
  & HD 188753 & $\gamma$Cep & HD 41004A & HD41004B & HD 196885 & $\alpha$CenB\\ \hline\hline
N\sub{DI}  &  0.54 &  1.41 &  1.57 &  1.25 &  1.45 &  1.15 \\
N\sub{CA}  & -2.19 &  7.04 &  6.87 &  3.83 &  7.13 &  3.45 \\
\hline
\end{tabular}

\end{center}
\end{table*}

The values of $N\sub{DI}$ for all binary systems considered 
are less than 2, as shown both in Table \ref{tab:binaries} 
and Table \ref{tab:checkstars}.  
This could be an indication that core accretion is 
more likely than disk instability in close binaries. 
As shown in Figure \ref{fig:diskstruct}, the disk model 
with the lowest $Q$ values has the highest accretion rate,  
$10^{-4}\;M_{\odot}\mbox{ yr}^{-1}$ and 
the lowest viscosity parameter explored.  
Disks with this high an accretion rate are typically 
outburst systems, such as FU Orionis objects.  Accretion rates 
this high are typically transient.  The disk 
contains 100 $M_J$ or 0.1 $M_{\odot}$ within just 1 AU, so it 
is extremely massive, hence it is the most gravitationally 
unstable.  Because a steady-state disk with these properties 
is unlikely, planet formation via disk instability is also unlikely.  

If we were to set a minimum threshold for $N\sub{CA}$ for 
which planet formation can occur in a close binary given the 
binary systems explored in this paper, we first need to consider 
which systems to include.  The existence of $\alpha$ Cen Bb has 
yet to be definitively confirmed, and it has the lowest 
value of $N\sub{CA}$ of 3, not including HD 188753A.  
If $\alpha$ Cen Bb is confirmed, then we might set 
$N\sub{CA,min}=3$.  
One might dispute that HD 41004Bb is a brown dwarf and ought 
not be considered, also.  Then, we might set $N\sub{CA,min}=7$, 
since $\gamma$ Cep A, HD 41004A, and HD 196885A all satisfy 
this criterion.  
Setting $N\sub{CA,min}=7$ is the more stringent constraint 
for planet formation, since we require more disks to satisfy 
the requirement for sufficient solid mass.
Based on these considerations, we 
predict that $N\sub{CA}\gtrsim7$ for planet formation 
to be readily feasible, based on the known planets in binary systems.  
For $3<N\sub{CA}<7$, planet formation may still be possible,
as in $\alpha$ Cen A, but less common.

\begin{figure*}
\includegraphics[width=\columnwidth]{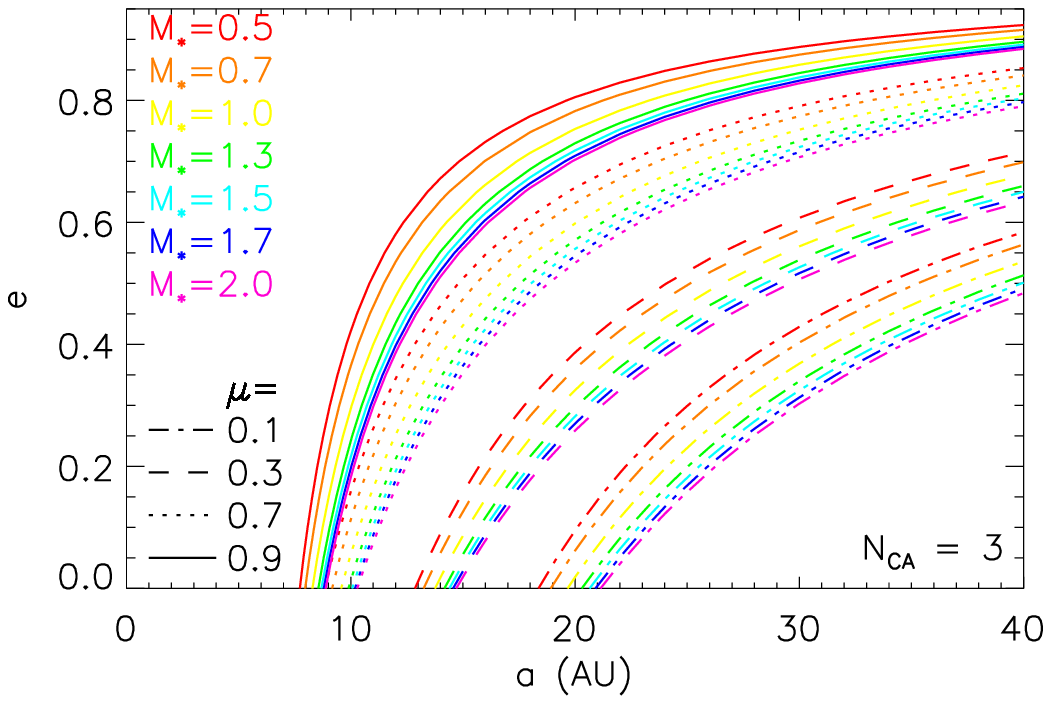}
\includegraphics[width=\columnwidth]{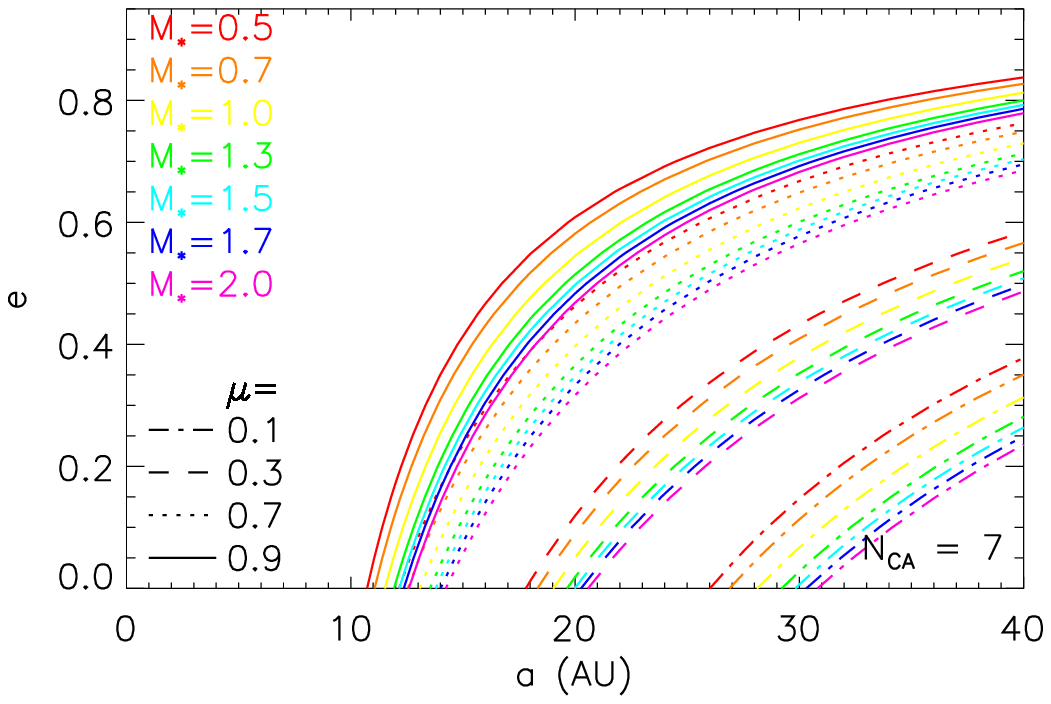}
\caption{\label{fig:CAcut}
Contours in $a$-$e$ space for $N\sub{CA}=3$ 
(left) and $N\sub{CA}=7$ (right)
for varying binary masses.  
Each contour represents a selected choice for $M_*$ and $\mu$, and the 
axes show allowed values of $a$ and $e$.  
The mass of the primary star is labeled by color, and the mass ratio 
of its binary companion is indicated by line style: 
dot-dashed, dashed, dotted, and solid for $\mu=0.1, 0.3, 0.7,$ and $0.9$, 
respectively.  Planet formation becomes more feasible 
to the right and below each curve.  
}
\end{figure*}

In Figure \ref{fig:CAcut}, we show the allowed parameter space 
for planet formation in close binaries, with 
$N\sub{CA,min}=3$ (left) or $N\sub{CA,min}=7$ (right).  
That is, we set equation (\ref{coeffs}) equal to 
3 or 7, choose fixed values of $M_*$ and $\mu$, and 
draw the resulting curve in $a$-$e$ space.  
Since equation (\ref{coeffs}) is linear in $a$ and $e$,
the resulting curve for $N=N\sub{CA,min}$ is 
hyperbolic in $a$-$e$ space for fixed $M_*$ and $\mu$.  

For a given binary pair, giant 
planet formation by core accretion 
is allowed for greater separations and lower eccentricity 
than the indicated contour, or under each curve.  
Planet formation also becomes more feasible as the primary mass decreases 
and the mass ratio increases, as shown by the 
shifting of the contours up and to the left.  

Note that for $a=12$ AU, the semi-major axis of the HD 188753 binary, 
planet formation is highly disfavored.  Even considering 
$N\sub{CA,min}=3$, only stars with relatively low-mass companions 
($\mu\gtrsim0.7$ or $M_b/M_*\lesssim3/7$) and low eccentricity 
($e\lesssim0.5$) might allow planet formation to occur.  
However, for $a=20$ AU, a large range of binary properties 
allow for planet formation to occur, although larger values of $\mu$
are favored.

\section{Discussion and Conclusions}

We have calculated the feasibility of giant planet formation in the close 
binary systems HD 188753, $\gamma$ Cep, HD 41004, 
HD 196885, and $\alpha$ Cen.  
We find that except for HD 188753, where there is not a planet, 
core accretion is the more likely 
planet formation mechanism than disk instability based on simple 
models of the truncated protoplanetary disk from which the planet 
formed.  These truncated disks contain 
a sufficiency of solid material from 
which planet cores can form, but do not meet the criterion for 
gravitational instability.  Based on the results for these 
systems, we conclude that for binary systems of separations 
$\lesssim$20 AU and eccentricity $\gtrsim$0.4, circumstellar 
giant planets may only form via core accretion.

More generally, we extend our results to binary systems of 
arbitrary masses, separations, and eccentricities to 
predict the likelihood of giant planet formation by either 
core accretion or disk instability.  These likelihoods are represented 
by the quantities $N\sub{CA}$ and $N\sub{DI}$, respectively.  
We find an empirical fit to $N\sub{CA}$ and $N\sub{DI}$ 
as a function of binary mass, mass ratio, separation.  
By comparing $N\sub{CA}$ and $N\sub{DI}$ as predicted 
in this paper to the observed frequency of planets in binaries, 
we can gain insight into 
the planet formation process in general.  

The estimates of planet formation likelihood represented by 
$N\sub{CA}$ and $N\sub{CI}$ are not meant to be definitive 
relations for the probability of planet formation.  
Rather, these values should be used to represent the 
relative ease with which planet formation can take 
place in binary systems.  As more planets are found in close 
binary systems, those statistics can be compared to 
the likelihoods calculated in this work to yield a better 
understanding of how planets can form in the dynamically 
challenging environment of a close binary star.  

We have considered only giant planet formation.  In principle, 
the requirement for terrestrial planet formation is less stringent, 
since less solid material is required.  
These calculations would be a possible direction for future work.  

Our results have been calculated under the assumption that 
the circumstellar disk is coplanar with the binary orbit.  
It is conceivable that the disk could be misaligned with the 
binary, as is the case for the wide pre-main sequence binary 
HK Tau \citep{2014JensenAkeson}.  However, 
tidal effects from close stellar companions 
with $\lesssim20$ AU separations will probably cause the 
disk to become aligned.  
The orbits of circumbinary planets discovered by the Kepler mission 
are closely aligned with those of the binary, suggesting that 
disks are likely to be aligned with binary orbits 
\citep{2011Doyle+,2012Welsh+,2012Orosz+a,2012Orosz+b}.

We have made the assumption in the core accretion case that likelihood
of planet formation depends only on the total 
solid mass being above a certain threshold, 
and that it does not depend on the amount above that threshold.  
We do not calculate whether or not these solids can come together to
form a single core, nor do we consider the possibility of the
formation of multiple cores.
These types of calculation would involve detailed modeling of the dynamics 
of the particles within the disk, and is beyond the scope of this paper.

We also do not consider the growth and evolution of giant planets
within the disk beyond the initial formation. Dynamical effects such
as tidal torques on planetesimals and planet embryos, planet-planet
scattering, and disk migration will affect the survivability of
planets within the system.  These effects have been considered in some
individual systems by others \citep[e.g.][]{2011Thebault}.

We have not considered how metallicity or chemical composition of the
disk might affect our results.  Increased metallicity will affect both
the disk opacities and the amount of solids in the disks.
Qualitatively speaking, increased opacity will result in slightly
higher disk temperatures, which will decrease the refractory content
in the most severely truncated disks.  On the other hand, higher
metallicity will raise the refractory content overall, so these
effects will compete with each other.  This could result in a lower
likelihood of giant planet formation in the most tightly bound binary
systems, but increased likelihood in wider systems.

\acknowledgments

Thanks go to Eric Jensen and an anonymous referee for helpful comments 
that greatly improved this paper.  The author acknowledges support from 
NASA ATP Grant NNX12AD43G.  

\bibliographystyle{apj}
\bibliography{apj-jour,$HOME/PaRTY/Papers/planets,$HOME/PaRTY/Papers/jang-condell,../binaries}

\begin{thebibliography}{30}
\expandafter\ifx\csname natexlab\endcsname\relax\def\natexlab#1{#1}\fi

\bibitem[{{Artymowicz} \& {Lubow}(1994)}]{1994ArtymowiczLubow}
{Artymowicz}, P. \& {Lubow}, S.~H. 1994, \apj, 421, 651

\bibitem[{{Beckwith} {et~al.}(1990){Beckwith}, {Sargent}, {Chini}, \&
  {Guesten}}]{1990Beckwith+}
{Beckwith}, S.~V.~W., {Sargent}, A.~I., {Chini}, R.~S., \& {Guesten}, R. 1990,
  \aj, 99, 924

\bibitem[{{Chauvin} {et~al.}(2011){Chauvin}, {Beust}, {Lagrange}, \&
  {Eggenberger}}]{2011Chauvin_etal}
{Chauvin}, G., {Beust}, H., {Lagrange}, A.-M., \& {Eggenberger}, A. 2011, \aap,
  528, A8

\bibitem[{{Correia} {et~al.}(2008){Correia}, {Udry}, {Mayor}, {Eggenberger},
  {Naef}, {Beuzit}, {Perrier}, {Queloz}, {Sivan}, {Pepe}, {Santos}, \&
  {S{\'e}gransan}}]{2008Correia_etal}
{Correia}, A.~C.~M., {Udry}, S., {Mayor}, M., {Eggenberger}, A., {Naef}, D.,
  {Beuzit}, J.-L., {Perrier}, C., {Queloz}, D., {Sivan}, J.-P., {Pepe}, F.,
  {Santos}, N.~C., \& {S{\'e}gransan}, D. 2008, \aap, 479, 271

\bibitem[{{Doyle} {et~al.}(2011){Doyle}, {Carter}, {Fabrycky}, {Slawson},
  {Howell}, {Winn}, {Orosz}, {Pr\v{s}a}, {Welsh}, {Quinn}, {Latham}, {Torres},
  {Buchhave}, {Marcy}, {Fortney}, {Shporer}, {Ford}, {Lissauer}, {Ragozzine},
  {Rucker}, {Batalha}, {Jenkins}, {Borucki}, {Koch}, {Middour}, {Hall},
  {McCauliff}, {Fanelli}, {Quintana}, {Holman}, {Caldwell}, {Still},
  {Stefanik}, {Brown}, {Esquerdo}, {Tang}, {Furesz}, {Geary}, {Berlind},
  {Calkins}, {Short}, {Steffen}, {Sasselov}, {Dunham}, {Cochran}, {Boss},
  {Haas}, {Buzasi}, \& {Fischer}}]{2011Doyle+}
{Doyle}, L.~R., {Carter}, J.~A., {Fabrycky}, D.~C., {Slawson}, R.~W., {Howell},
  S.~B., {Winn}, J.~N., {Orosz}, J.~A., {Pr\v{s}a}, A., {Welsh}, W.~F.,
  {Quinn}, S.~N., {Latham}, D., {Torres}, G., {Buchhave}, L.~A., {Marcy},
  G.~W., {Fortney}, J.~J., {Shporer}, A., {Ford}, E.~B., {Lissauer}, J.~J.,
  {Ragozzine}, D., {Rucker}, M., {Batalha}, N., {Jenkins}, J.~M., {Borucki},
  W.~J., {Koch}, D., {Middour}, C.~K., {Hall}, J.~R., {McCauliff}, S.,
  {Fanelli}, M.~N., {Quintana}, E.~V., {Holman}, M.~J., {Caldwell}, D.~A.,
  {Still}, M., {Stefanik}, R.~P., {Brown}, W.~R., {Esquerdo}, G.~A., {Tang},
  S., {Furesz}, G., {Geary}, J.~C., {Berlind}, P., {Calkins}, M.~L., {Short},
  D.~R., {Steffen}, J.~H., {Sasselov}, D., {Dunham}, E.~W., {Cochran}, W.~D.,
  {Boss}, A., {Haas}, M.~R., {Buzasi}, D., \& {Fischer}, D. 2011, Science, 333,
  1602

\bibitem[{{Dumusque} {et~al.}(2012){Dumusque}, {Pepe}, {Lovis},
  {S{\'e}gransan}, {Sahlmann}, {Benz}, {Bouchy}, {Mayor}, {Queloz}, {Santos},
  \& {Udry}}]{2012Dumusque_etal}
{Dumusque}, X., {Pepe}, F., {Lovis}, C., {S{\'e}gransan}, D., {Sahlmann}, J.,
  {Benz}, W., {Bouchy}, F., {Mayor}, M., {Queloz}, D., {Santos}, N., \& {Udry},
  S. 2012, \nat, 491, 207

\bibitem[{{Eggenberger} {et~al.}(2007){Eggenberger}, {Udry}, {Mazeh}, {Segal},
  \& {Mayor}}]{2007Eggenberger_etal}
{Eggenberger}, A., {Udry}, S., {Mazeh}, T., {Segal}, Y., \& {Mayor}, M. 2007,
  \aap, 466, 1179

\bibitem[{{Endl} {et~al.}(2011){Endl}, {Cochran}, {Hatzes}, \&
  {Wittenmyer}}]{2011Endl_etal}
{Endl}, M., {Cochran}, W.~D., {Hatzes}, A.~P., \& {Wittenmyer}, R.~A. 2011, in
  American Institute of Physics Conference Series, Vol. 1331, American
  Institute of Physics Conference Series, ed. S.~{Schuh}, H.~{Drechsel}, \&
  U.~{Heber}, 88--94

\bibitem[{{Fischer} {et~al.}(2009){Fischer}, {Driscoll}, {Isaacson}, {Giguere},
  {Marcy}, {Valenti}, {Wright}, {Henry}, {Johnson}, {Howard}, {Peek}, \&
  {McCarthy}}]{2009Fischer_etal}
{Fischer}, D., {Driscoll}, P., {Isaacson}, H., {Giguere}, M., {Marcy}, G.~W.,
  {Valenti}, J., {Wright}, J.~T., {Henry}, G.~W., {Johnson}, J.~A., {Howard},
  A., {Peek}, K., \& {McCarthy}, C. 2009, \apj, 703, 1545

\bibitem[{{Hatzes}(2013)}]{2013Hatzes}
{Hatzes}, A.~P. 2013, \apj, 770, 133

\bibitem[{{Hatzes} {et~al.}(2003){Hatzes}, {Cochran}, {Endl}, {McArthur},
  {Paulson}, {Walker}, {Campbell}, \& {Yang}}]{2003Hatzes_etal}
{Hatzes}, A.~P., {Cochran}, W.~D., {Endl}, M., {McArthur}, B., {Paulson},
  D.~B., {Walker}, G.~A.~H., {Campbell}, B., \& {Yang}, S. 2003, \apj, 599,
  1383

\bibitem[{{Holman} \& {Wiegert}(1999)}]{1999HolmanWiegert}
{Holman}, M.~J. \& {Wiegert}, P.~A. 1999, \aj, 117, 621

\bibitem[{{Jang-Condell}(2007)}]{HJChd188753}
{Jang-Condell}, H. 2007, \apj, 654, 641

\bibitem[{{Jang-Condell} {et~al.}(2008){Jang-Condell}, {Mugrauer}, \&
  {Schmidt}}]{2008HJCMugrauerSchmidt}
{Jang-Condell}, H., {Mugrauer}, M., \& {Schmidt}, T. 2008, \apjl, 683, L191

\bibitem[{{Jensen} \& {Akeson}(2014)}]{2014JensenAkeson}
{Jensen}, E.~L.~N. \& {Akeson}, R. 2014, \nat, 511, 567

\bibitem[{{Jensen} {et~al.}(1994){Jensen}, {Mathieu}, \&
  {Fuller}}]{1994Jensen+}
{Jensen}, E.~L.~N., {Mathieu}, R.~D., \& {Fuller}, G.~A. 1994, \apjl, 429, L29

\bibitem[{{Jensen} {et~al.}(1996){Jensen}, {Mathieu}, \&
  {Fuller}}]{1996Jensen+}
---. 1996, \apj, 458, 312

\bibitem[{{Konacki}(2005)}]{HD188753}
{Konacki}, M. 2005, Nature, 436, 230

\bibitem[{{Orosz} {et~al.}(2012{\natexlab{a}}){Orosz}, {Welsh}, {Carter},
  {Brugamyer}, {Buchhave}, {Cochran}, {Endl}, {Ford}, {MacQueen}, {Short},
  {Torres}, {Windmiller}, {Agol}, {Barclay}, {Caldwell}, {Clarke}, {Doyle},
  {Fabrycky}, {Geary}, {Haghighipour}, {Holman}, {Ibrahim}, {Jenkins},
  {Kinemuchi}, {Li}, {Lissauer}, {Pr{\v s}a}, {Ragozzine}, {Shporer}, {Still},
  \& {Wade}}]{2012Orosz+b}
{Orosz}, J.~A., {Welsh}, W.~F., {Carter}, J.~A., {Brugamyer}, E., {Buchhave},
  L.~A., {Cochran}, W.~D., {Endl}, M., {Ford}, E.~B., {MacQueen}, P., {Short},
  D.~R., {Torres}, G., {Windmiller}, G., {Agol}, E., {Barclay}, T., {Caldwell},
  D.~A., {Clarke}, B.~D., {Doyle}, L.~R., {Fabrycky}, D.~C., {Geary}, J.~C.,
  {Haghighipour}, N., {Holman}, M.~J., {Ibrahim}, K.~A., {Jenkins}, J.~M.,
  {Kinemuchi}, K., {Li}, J., {Lissauer}, J.~J., {Pr{\v s}a}, A., {Ragozzine},
  D., {Shporer}, A., {Still}, M., \& {Wade}, R.~A. 2012{\natexlab{a}}, \apj,
  758, 87

\bibitem[{{Orosz} {et~al.}(2012{\natexlab{b}}){Orosz}, {Welsh}, {Carter},
  {Fabrycky}, {Cochran}, {Endl}, {Ford}, {Haghighipour}, {MacQueen}, {Mazeh},
  {Sanchis-Ojeda}, {Short}, {Torres}, {Agol}, {Buchhave}, {Doyle}, {Isaacson},
  {Lissauer}, {Marcy}, {Shporer}, {Windmiller}, {Barclay}, {Boss}, {Clarke},
  {Fortney}, {Geary}, {Holman}, {Huber}, {Jenkins}, {Kinemuchi}, {Kruse},
  {Ragozzine}, {Sasselov}, {Still}, {Tenenbaum}, {Uddin}, {Winn}, {Koch}, \&
  {Borucki}}]{2012Orosz+a}
{Orosz}, J.~A., {Welsh}, W.~F., {Carter}, J.~A., {Fabrycky}, D.~C., {Cochran},
  W.~D., {Endl}, M., {Ford}, E.~B., {Haghighipour}, N., {MacQueen}, P.~J.,
  {Mazeh}, T., {Sanchis-Ojeda}, R., {Short}, D.~R., {Torres}, G., {Agol}, E.,
  {Buchhave}, L.~A., {Doyle}, L.~R., {Isaacson}, H., {Lissauer}, J.~J.,
  {Marcy}, G.~W., {Shporer}, A., {Windmiller}, G., {Barclay}, T., {Boss},
  A.~P., {Clarke}, B.~D., {Fortney}, J., {Geary}, J.~C., {Holman}, M.~J.,
  {Huber}, D., {Jenkins}, J.~M., {Kinemuchi}, K., {Kruse}, E., {Ragozzine}, D.,
  {Sasselov}, D., {Still}, M., {Tenenbaum}, P., {Uddin}, K., {Winn}, J.~N.,
  {Koch}, D.~G., \& {Borucki}, W.~J. 2012{\natexlab{b}}, Science, 337, 1511

\bibitem[{{Osterloh} \& {Beckwith}(1995)}]{1995OsterlohBeckwith}
{Osterloh}, M. \& {Beckwith}, S.~V.~W. 1995, \apj, 439, 288

\bibitem[{{Pourbaix} {et~al.}(2002){Pourbaix}, {Nidever}, {McCarthy}, {Butler},
  {Tinney}, {Marcy}, {Jones}, {Penny}, {Carter}, {Bouchy}, {Pepe}, {Hearnshaw},
  {Skuljan}, {Ramm}, \& {Kent}}]{2002Pourbaix_etal}
{Pourbaix}, D., {Nidever}, D., {McCarthy}, C., {Butler}, R.~P., {Tinney},
  C.~G., {Marcy}, G.~W., {Jones}, H.~R.~A., {Penny}, A.~J., {Carter}, B.~D.,
  {Bouchy}, F., {Pepe}, F., {Hearnshaw}, J.~B., {Skuljan}, J., {Ramm}, D., \&
  {Kent}, D. 2002, \aap, 386, 280

\bibitem[{{Pringle}(1981)}]{pringle}
{Pringle}, J.~E. 1981, \araa, 19, 137

\bibitem[{{Raghavan} {et~al.}(2006){Raghavan}, {Henry}, {Mason}, {Subasavage},
  {Jao}, {Beaulieu}, \& {Hambly}}]{2006Raghavan_etal}
{Raghavan}, D., {Henry}, T.~J., {Mason}, B.~D., {Subasavage}, J.~P., {Jao},
  W.-C., {Beaulieu}, T.~D., \& {Hambly}, N.~C. 2006, \apj, 646, 523

\bibitem[{{Shakura} \& {Sunyaev}(1973)}]{shaksun}
{Shakura}, N.~I. \& {Sunyaev}, R.~A. 1973, \aap, 24, 337

\bibitem[{{Siess} {et~al.}(2000){Siess}, {Dufour}, \& {Forestini}}]{siess_etal}
{Siess}, L., {Dufour}, E., \& {Forestini}, M. 2000, \aap, 358, 593

\bibitem[{{Thebault}(2011)}]{2011Thebault}
{Thebault}, P. 2011, Celestial Mechanics and Dynamical Astronomy, 111, 29

\bibitem[{{Welsh} {et~al.}(2012){Welsh}, {Orosz}, {Carter}, {Fabrycky}, {Ford},
  {Lissauer}, {Pr{\v s}a}, {Quinn}, {Ragozzine}, {Short}, {Torres}, {Winn},
  {Doyle}, {Barclay}, {Batalha}, {Bloemen}, {Brugamyer}, {Buchhave},
  {Caldwell}, {Caldwell}, {Christiansen}, {Ciardi}, {Cochran}, {Endl},
  {Fortney}, {Gautier}, {Gilliland}, {Haas}, {Hall}, {Holman}, {Howard},
  {Howell}, {Isaacson}, {Jenkins}, {Klaus}, {Latham}, {Li}, {Marcy}, {Mazeh},
  {Quintana}, {Robertson}, {Shporer}, {Steffen}, {Windmiller}, {Koch}, \&
  {Borucki}}]{2012Welsh+}
{Welsh}, W.~F., {Orosz}, J.~A., {Carter}, J.~A., {Fabrycky}, D.~C., {Ford},
  E.~B., {Lissauer}, J.~J., {Pr{\v s}a}, A., {Quinn}, S.~N., {Ragozzine}, D.,
  {Short}, D.~R., {Torres}, G., {Winn}, J.~N., {Doyle}, L.~R., {Barclay}, T.,
  {Batalha}, N., {Bloemen}, S., {Brugamyer}, E., {Buchhave}, L.~A., {Caldwell},
  C., {Caldwell}, D.~A., {Christiansen}, J.~L., {Ciardi}, D.~R., {Cochran},
  W.~D., {Endl}, M., {Fortney}, J.~J., {Gautier}, III, T.~N., {Gilliland},
  R.~L., {Haas}, M.~R., {Hall}, J.~R., {Holman}, M.~J., {Howard}, A.~W.,
  {Howell}, S.~B., {Isaacson}, H., {Jenkins}, J.~M., {Klaus}, T.~C., {Latham},
  D.~W., {Li}, J., {Marcy}, G.~W., {Mazeh}, T., {Quintana}, E.~V., {Robertson},
  P., {Shporer}, A., {Steffen}, J.~H., {Windmiller}, G., {Koch}, D.~G., \&
  {Borucki}, W.~J. 2012, \nat, 481, 475

\bibitem[{{Zucker} {et~al.}(2003){Zucker}, {Mazeh}, {Santos}, {Udry}, \&
  {Mayor}}]{2003Zucker_HD41004}
{Zucker}, S., {Mazeh}, T., {Santos}, N.~C., {Udry}, S., \& {Mayor}, M. 2003,
  \aap, 404, 775

\bibitem[{{Zucker} {et~al.}(2004){Zucker}, {Mazeh}, {Santos}, {Udry}, \&
  {Mayor}}]{2004Zucker_HD41004}
---. 2004, \aap, 426, 695

\end{thebibliography}

\end{document}